# A Trick: Why $\hat{\gamma} < \gamma$ in [1]?

Wuhua Hu

*School of Electrical Engineering and Automation, Tianjin University, Tianjin 300072, P. R. China*

*Email: haitun_ahua@yahoo.com.cn*

***Abstract:*** In this paper, first a theorem on the partial sum of a particular series is given. Then, based on it, the origin of obvious simulation deviation from theory is explained: *i)* why the numerically estimated $\hat{\gamma}$ (degree exponent) in [1] is always smaller than $\gamma \equiv 3$ that is predicted by theory; *ii)* and why $\hat{\gamma}$ rises monotonically as $m$ (the links added at each step in Barabási-Albert (BA) model [1]) increases. Strictly, it declares such errors are basically from the inconsistence of simulation with the theoretical model, which is caused by an additional incompatible condition used in simulation. In addition, noticing the evolving differences between the initial $m_0$ nodes and those after, we correct the derived BA model which unfairly omitted such differences.

If look carefully, we will find that some flaw exists in Part B, Section VII of paper [1]: the simulation results are in imperfect consistence with those derived by their theory. Looking at Fig. 1 (a copy of FIG. 21 in [1]), we can find that the absolute slopes of the plots, with more or less $\hat{\gamma} = 2.9$ [1], are all smaller than $\gamma \equiv 3$, as predicted in theory, and also that $\hat{\gamma}$ rises as $m$ increases.

The authors omit these errors, though. Without any particular explanation, they might have attributed them to the inaccurate numerical calculation or inexact theoretical result. But, we would more like to believe in the exactness of its theory. Then, a question comes up: What is wrong with the numerical method, i.e., where are the errors from? Are they the common kinds of errors in numerical calculation, such as the rounding error? They seem not so. Now, here we are trying to uncover this puzzle. First, let us see one theorem.

**Theorem 1.** For any $1 < n \in Z^+, p \in R^+$, let

$$S_n =: \sum_{i=1}^{n} \frac{1}{i^p}, \quad S_{n,1} =: \frac{(n+1)^{1-p} - 1}{1-p}, \quad S_{n,2} =: \frac{n^{1-p} - p}{1-p}, \quad (1)$$
$$E_{n,1} =: S_n - S_{n,1}, \quad E_{n,2} =: S_{n,2} - S_n.$$

There hold *i)* $S_{n,1} < S_n < S_{n,2}$; *ii)* $E_{n,1}$ and $E_{n,2}$ are increasing in $n$. □

*Proof.* "*i)*"

$$\left.\begin{array}{l} S_n = \sum_{i=1}^{n} \frac{1}{i^p} \int_{i-1}^{i} dx \\ i-1 \leq x \leq i \\ \frac{\partial (1/x^p)}{\partial x} < 0 \end{array}\right\} \Rightarrow \sum_{i=1}^{n} \int_{i-1}^{i} \frac{1}{(x+1)^p} dx < S_n < 1 + \sum_{i=2}^{n} \int_{i-1}^{i} \frac{1}{x^p} dx$$

$$\Rightarrow \frac{(n+1)^{1-p} - 1}{1-p} = S_{n,1} < S_n < S_{n,2} = \frac{n^{1-p} - p}{1-p}.$$

"*ii)*"

$$E_{n+1,2} - E_{n,2} = \left( \frac{(n+1)^{1-p} - p}{1-p} - \sum_{i=1}^{n+1} \frac{1}{i^p} \right) - \left( \frac{n^{1-p} - p}{1-p} - \sum_{i=1}^{n} \frac{1}{i^p} \right)$$
$$= \frac{(n+1)^{1-p} - n^{1-p}}{1-p} - \frac{1}{(n+1)^p}$$
$$= \int_{n}^{n+1} \frac{1}{x^p} dx - \frac{1}{(n+1)^p} > 0.$$

So, $E_{n,2}$ is increasing in $n$. And similarly, we can prove that $E_{n,1}$ is increasing in $n$, too. □

Thus, taking $p = 1/2$ in (1), we obtain

$$2(\sqrt{n+1} - 1) < S_n =: \sum_{i=1}^{n} \frac{1}{\sqrt{i}} < 2\sqrt{n} - 1, \quad (2)$$
and $E_{n,2} =: 2\sqrt{n} - 1 - S_n$ is increasing in *n*.

Now, let us consider the BA model derived by continuum theory presented in [1] and make some necessary modification on it. When supposing that the initial $m_0$ nodes are unlinked, the way of taking the time span as $[0, m_0 + t]$ is improper indeed, because the degrees of the first $m_0$ nodes (i.e., $t \in [0, m_0]$) do not obey the model of $k_i(t) = m(t/t_i)^\beta$, and instead they keep zero unless $t > m_0$.

Besides, the initial $m_0$ nodes cannot meet the condition $k_i(t_i) = m$, as used in [1]. Considering these, the model needs modification. And it can be done like this: *a)* Take the time when the initial $m_0$ nodes exist as zero; *b)* Take $m_0 \equiv m$, so that $k_i(t_1) \equiv 1$ for $i = 1, 2, \ldots, m_0$ (the initial $m_0$ nodes); *c)* The upper bound of $j$ in (3) should be $m_0 + t$, rather than $m_0 + t - 1$ as meant in [1], because (3) means the variation rate of $k_i$ at time $t$, rather than at time $t-1$. Then, using the continuum theory, the BA model can be derived as follows.

$$\frac{\partial k_i}{\partial t} = m\Pi(k_i) = m \frac{k_i}{\sum_{j=1}^{m_0+t} k_j}. \tag{3}$$

With the initial condition,

$$\begin{cases} k_i(t_1) = 1, \ i = 1, 2, \ldots, m_0 \\ k_i(t_{i-m_0}) = m, \ i = m_0 + 1, m_0 + 2, \ldots, m_0 + t \end{cases} \tag{4}$$

we can obtain

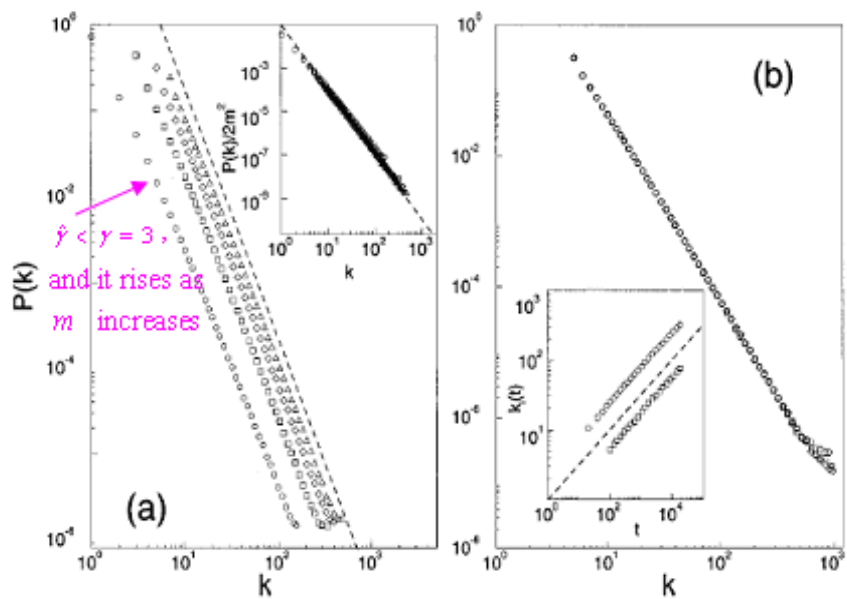

Fig.1 ([1]) Numerical simulations of network evolution: (a) Degree distribution of the Barabási-Albert model, with $N = m_0 + t = 300\,000$ and O, $m_0 = m = 1$; □, $m_0 = m = 3$; ◊, $m_0 = m = 5$; and △, $m_0 = m = 7$. The slope of the dashed line is $\gamma = 2.9$, providing the best fit to the data. The inset shows the rescaled distribution (see text) $P(k)/2m^2$ for the same values of $m$, the slope of the dashed line being $\gamma = 3$; (b) $P(k)$ for $m_0 = m = 5$ and various system sizes, O, $N = 100\,000$; □, $N = 150\,000$; ◊, $N = 200\,000$. The inset shows the time evolution for the degree of two vertices, added to the system at $t_1 = 5$ and $t_2 = 95$. Here $m_0 = m = 5$, and the dashed line has slope 0.5, as predicted by Eq. (81). After Barabási, Albert, and Jeong (1999).

$$k_i(t) = \begin{cases} \left(\dfrac{t}{t_1}\right)^\beta, & i=1,\ 2,\ \ldots,\ m_0 \\ m\left(\dfrac{t}{t_{i-m_0}}\right)^\beta, & i=m_0+1,\ m_0+2,\ \ldots,\ m_0+t \end{cases},\ \text{with } \beta=\dfrac{1}{2}. \tag{5}$$

Then, similar to [1], it has $\gamma = 1 + 1/\beta = 3$. However, it should be noticed that one condition must be satisfied during the deduction, namely

$$\sum_{i=1}^{m_0+t} k_i(t) = 2mt. \tag{6}$$

Let

$$S_t =: \sum_{i=1}^{m_0} \frac{1}{m\sqrt{t_1}} + \sum_{i=m_0+1}^{m_0+t} \frac{1}{\sqrt{t_{i-m_0}}} = \frac{1}{\sqrt{t_1}} + \sum_{i=1}^{t} \frac{1}{\sqrt{t_i}} \stackrel{\text{if } t_1=1}{=} \sum_{i=1}^{t} \frac{1}{\sqrt{t_i}} + 1. \tag{7}$$

Substitute (5) into (6), and we obtain

$$S_t = 2t^{1-\beta} = 2\sqrt{t} \iff \sum_{i=1}^{t} \frac{1}{\sqrt{t_i}} = S_t - \frac{1}{\sqrt{t_1}} \stackrel{\text{if } t_1=1}{=} 2\sqrt{t} - 1. \tag{8}$$

So it means that only if (8) stands can (5) stand. But, if we take

$$t_i = i,\ \text{where}\ i=1,\ 2,\ \ldots,\ t, \tag{9}$$

can (8) still hold? Indeed, that needs a careful check. According to (2), for any $t > 1$, there holds

$$2(\sqrt{t+1}-1) < \sum_{i=1}^{t} \frac{1}{\sqrt{t_i}} = \sum_{i=1}^{t} \frac{1}{\sqrt{i}} < 2\sqrt{t}-1. \tag{10}$$

Hence, (8) does not stand in this case. And in fact, under the condition of (9),

$$S_t = 2t^{1-\hat{\beta}} < 2\sqrt{t} = 2t^{1-\beta} \iff \hat{\beta} > \beta = 1/2 \iff \hat{\gamma} = 1+1/\hat{\beta} < \gamma = 1+1/\beta = 3, \tag{11}$$

where $\hat{\beta}$ is the numerically estimative value of $\beta$. (11) means that if we used the numerical value $\hat{\gamma}$ to estimate the true value $\gamma$, it should be smaller. So, this successfully explains the phenomena that the absolute slopes of the plots in Fig. 1. (a) are all smaller than $\gamma \equiv 3$ in theory.

Besides, according to (2), (7), (9) and (11), there holds

$$E_{t,2} := 2\sqrt{t} - 1 - \sum_{i=1}^{t} \frac{1}{\sqrt{i}} = 2\sqrt{t} - S_t = 2(t^{1/2} - t^{1-\hat{\beta}}) > 0 \text{ is increasing in } t. \tag{12}$$

If we set $m_0 + t = m + t = const$ when deriving the numerical results, then $t$ would decrease monotonically as $m$ went up. And so would $E_{t,2}$, because of (12). Consequently, $t^{1-\hat{\beta}}$ would monotonically approach $t^{1/2}$ from the bottom. And hence, $\hat{\beta}$ would monotonically approach $1/2$ from the bottom.

As a result, $\hat{\gamma} = 1 + 1/\hat{\beta}$ rises monotonically as $m$ increases, and $\hat{\gamma} < \gamma = 3$ will always hold. Consequently, as $m$ increases, the plots become stiffer in Fig. 1. (a) and their absolute slopes are always smaller than 3.

**Conclusion**

First, noticing the difference of the initial $m_0$ nodes from those newly added ones, we modified the BA model derived in [1]. Second, noticing the necessary condition (8) for (3) to stand, we used Theorem 1 and demonstrated that: Because the additional condition (9) used in numerical simulation makes (8) unsatisfied and (11) hold, the numerically derived $\hat{\gamma}$ is doomed to be smaller than $\gamma$ that is obtained theoretically. What is more, when setting $m + t$ as a const, as $\hat{\gamma}$ is monotonically increasing with respect to $m$, the plots become absolutely stiffer as $m$ increases.

In a great sense, the theory presented here uncovers that the deviation of simulation from theoretical model in [1] is due to a simulation fault rather than the common numerical calculating errors. And, this generally warns of taking care to interpret the numerical results and to do simulations. Besides, Theorem 1 presented in this paper can give guidance for similar simulations in the future.